\documentclass[10pt]{elsart}
\usepackage{graphicx}
\usepackage{subfigure}
\usepackage{amsmath}
\usepackage{color}
\usepackage{amsfonts}
\usepackage{amssymb}

\begin{document}
\begin{frontmatter}
\title{Statistical analysis of simple repeats in the human genome}
\author[label1]{F. Piazza}
\ead{Francesco.Piazza@epfl.ch}
\author[label2]{P. Li\`o}
\ead{Pietro.Lio@cl.cam.ac.uk}
\address[label1]{Laboratoire de Biophysique Statistique, ITP-FSB,
\'Ecole Politechnique F\'ed\'erale de Lausanne,
CH-1015 Lausanne, Switzerland}
\address
[label2]{Computer Laboratory, University of Cambridge, Cambridge CB3 0FD, UK}
\begin{abstract}
The human genome contains repetitive DNA at different level of sequence
length, number and dispersion. Highly repetitive DNA is particularly rich in
homo-- and di--nucleotide repeats, while middle repetitive DNA is rich of
families of interspersed, mobile elements hundreds of base pairs (bp) long, among
which the Alu families.
A link between homo- and di-polymeric tracts and mobile elements has been recently
highlighted. In particular, the mobility of Alu repeats, which form 10\% of the
human genome, has been correlated with the length of poly(A) tracts
located at one end of the Alu. These tracts have a rigid and non-bendable
structure and have an inhibitory effect on nucleosomes, which normally
compact the DNA.
We performed a statistical analysis of the genome-wide
distribution of lengths and inter--tract separations of poly(X) and
poly(XY) tracts in the human genome. Our study shows that in humans the
length distributions of these sequences
reflect the  dynamics of their expansion and DNA replication.
By means of general tools from linguistics, we show that the latter play the role of
highly-significant content-bearing terms  in the DNA text.
Furthermore, we find that such tracts are positioned in a non-random fashion,
with an apparent periodicity of 150 bases.
This allows us to extend the link between repetitive, highly mobile elements such as
Alus and low-complexity words in human DNA.
More precisely, we show that Alus are sources of poly(X) tracts, which in turn
affect in a subtle way the combination and
diversification of gene expression and the fixation of multigene families.
\end{abstract}
\begin{keyword}
DNA \sep
Homopolymeric repeats (HR), Alu repeats, mobile elements, human genome
\PACS
87.10.+e; 02.50.Cw
\end{keyword}
\end{frontmatter}

%

\section{Introduction}

Experiments on kinetics of DNA denaturation and renaturation and the analysis
of DNA sequences have revealed that most of our genome is populated by DNA
repeats of different length, number and degree of dispersion~\cite{Lander2001}%
. Long repeats in few copies are usually orthologous genes, which may contain
hidden repeats in form of runs of amino acids, and retroviruses inserted in
the genome. For example, the human genome contains more than 50 chemokine
receptor genes which have high sequence similarity~\cite{chemokine2003} and
almost one thousands olfactory receptor genes and
pseudogenes~\cite{olfactory2001}. Short repetitive DNA sequences may be
categorized in highly and middle repetitive. The first is formed by tandemly
clustered DNA of variable length motifs (5-100 bp) and is present in large
islands of up to 100 Mb. The middle-repetitive can be either short islands of
tandemly repeated microsatellites/minisatellites (`CA repeats', tri- and
tetra-nucleotide repeats) or mobile genetic elements. Mobile elements include
DNA transposons, short and long interspersed elements (SINEs and LINEs), and
processed pseudogenes~\cite{Weiner2002,WLI}.

Why should we be interested in repetitive DNA? Tandem repeats with 1-3 base
motif can differ in repeat number among individuals; therefore, they are used
as genetic markers for assessing genetic differences in plants and animals and
in forensic testing. It is known that trinucleotide repeats are involved in
several human neurodegenerative diseases (e.g., fragile X and Huntington's
disease), and instability of short tandemly repeated DNA has been associated
with cancer~\cite{cancer,Bois1999,Gifford2004}. DNA repeats increase DNA
recombination events and have the potential to destroy (by insertional
mutagenesis), to create (by generating functional retropseudogenes), and to
empower (by giving old genes new promoters or regulatory signals). Despite
their importance in genome dynamics and for medical diagnosis, and despite the
advances in the understanding of their role in several prokaryotic and
eukaryotic genomes~\cite{Claudia,Holste2003,Mantegna1,Trivonov,Mantegna}, a
robust, genome--wide, statistical analysis of interspersed repetitive elements
in human genome is still lacking. In particular, the analysis of short
repetitive DNA needs to fully exploit the relationship between simple repeats
and mobile elements.

Interestingly, all mobile elements such as SINEs, LINEs, and processed
pseudogenes, contain A--rich regions of different length~\cite{Deininger2003}.
In particular, the Alu elements, present exclusively in the primates, are the
most abundant repeat elements in terms of copy number ($> 10^{6}$ in the human
genome) and account for more than 10\% of the human genome~\cite{Lander2001}.
They are typically 300 nucleotides in length, often form clusters, and are
mainly present in noncoding regions. Higher Alu densities were observed in
chromosomes with a greater number of genes and vice versa. Alus have a dimeric
structure and are ancestrally derived from the gene 7SL RNA. They amplify in
the genome by using a RNA polymerase III-derived transcript as template, in a
process termed retroposition~\cite{retroposition,Deininger2002}. The mobility
is facilitated by a variable-length stretch of an A-rich region located at the
3' end~\cite{Batzer2002,reverse}.

Although all Alus elements have poly(A) stretches, only a very few are able to
retropose~\cite{Deininger2003}. Therefore, the mere presence of a poly(A)
stretch is not sufficient to confer on an Alu element the ability to retropose
efficiently. However, the length of the A stretch correlates positively with
the mobility of the Alu~\cite{Roy-Engel2002}.

The Alu repeats are divided into three subfamilies on the basis of their
evolutionary age: Alu J (oldest), S (intermediate age) and Y
(youngest)~\cite{Johanning2003}. There is an inverse correlation between the
age of the Alu subfamily and the proportion of the members with long A--tails
in the genome, indicating that loss of A stretches may be a primary, though
not the only, inactivating feature in the older
subfamilies~\cite{Roy-Engel2002}.

In this study we first investigate exhaustively the distribution and
characteristic length size of all homopolymeric repeats (HR) of the kind
poly(X) in the complete human genome, where X $\in$ [A,C,G,T]. By means of
simple tools drawn from linguistics, we show that stretches of homopolymeric
repeats play a highly--specialized role in the DNA text. In addition, we show
that the former are more specific words within the human genome with respect
to other repeats coded from different alphabets (see Table~\ref{t:alphabets}
for a list of alphabets considered here).

We quantify this effect by studying the characteristic positioning patterns of
stretches of given composition and length. We then focus on long A stretches
in human chromosomes 20, 21 and 22. The latter chromosomes differ
substantially in both Alu density and gene density. Chromosomes 21 and 22, for
example, are of similar size (together about $1.6$ \% of the human genome),
even though chromosome 22 has four times as many genes and twice as many Alu
repeats~\cite{Karlin2002}. The comparative analysis of genome--wide
distribution of poly(A) and other homo--dinucleotide polymers allows us to
examine the mechanisms and constraints of poly(A) elongation or shortening,
and how the elongation dynamics is related to the evolutionary
instabilities~\cite{Weiner2002,Fryxell2000}, DNA
bendability~\cite{Bruckner1995}, and nucleosome inhibition~\cite{Englander}.

\section{Length distributions of poly(X) repeats in human genome}

For our study of homopolymeric tracts in the human genome, we used all the
finished sequences of the 24 chromosomes, among which the published sequences
of chromosomes 21 and 22, as well as a set of compiled sequences together
covering about 3 Giga bases (Gb). DNA sequences were obtained from the Genbank
directory of the web site of the National Center for Biotechnology Information
(\texttt{ftp://ncbi.nlm.nih.gov/}).

\begin{figure}[h]
\centering
\includegraphics[width=10 truecm,clip]{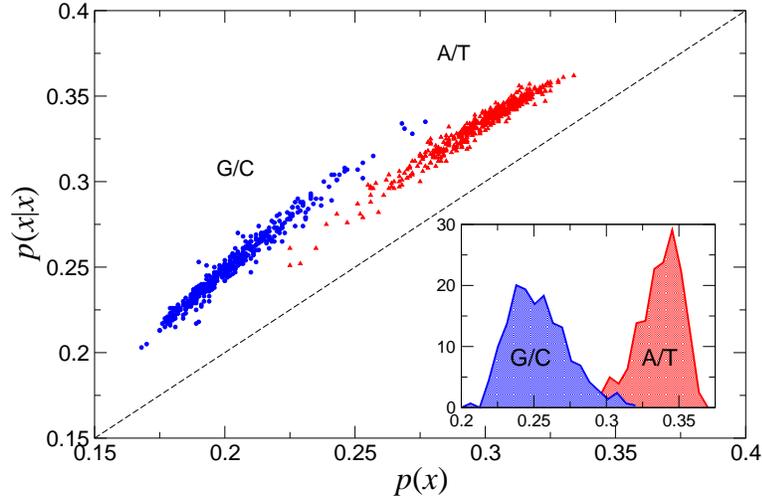}
\caption{Human Chromosomes 1, 3, 5 and 7. Plot of the conditional
probabilities $p(X|X)$ vs the single--nucleotide probabilities $p(X)$ for
X=A,T and X=C,G. The dashed line marks the totally uncorrelated (Bernoulli)
case $p(X|X)=p(X)$. Inset: normalized histograms of the conditional
probabilities. The probabilities are measured in windows of length 22 Kbp
along the chromosomes.}%
\label{f:pxx_vs_px}%
\end{figure}

We find that homopolymeric tracts of the type poly(X) are substantially
over--represented in all human chromosomes. A first simple quantitative
measure of over--abundance of poly(X) strings can be obtained by calculating
the conditional probabilities $p(X_{i+1}|X_{i})$, i.e. the probabilities of
finding nucleotide X at position $i+1$, given that position $i$ is also
occupied by X. In the totally uncorrelated case (sometimes referred to as the
Bernoulli case
%
%
) $p(X|X)=p(X)$, whereas $p(X|X)\neq p(X)$ denotes the presence of spatial
correlations along the sequence. This effect in the human genome is depicted
in Figure~\ref{f:pxx_vs_px}, where we show as an example data from chromosomes
1, 3, 5, and 7. The presence of positive correlations (here measured at the
di--nucleotide level) is very clear.

A more quantitative analysis of the correlations underlying the structure of
poly(X) words can be performed by studying the length distributions of such
repetitive sequences. For this purpose, it is better to work with the
cumulative (integrated) distributions, in order to gain some statistical
weight in the low--frequency regions. These can be easily obtained from the
data by noting that they are nothing but rank--size plots with the axes inverted.

We find that each chromosome displays approximately the same statistical
properties as for the length distributions of poly(X) strings, in both coding
and non--coding regions. In particular, the main features of such
distributions may be summarized as follows.

\begin{figure}[h]
\centering
\subfigure[]
{\includegraphics[width=6.7 truecm,clip]{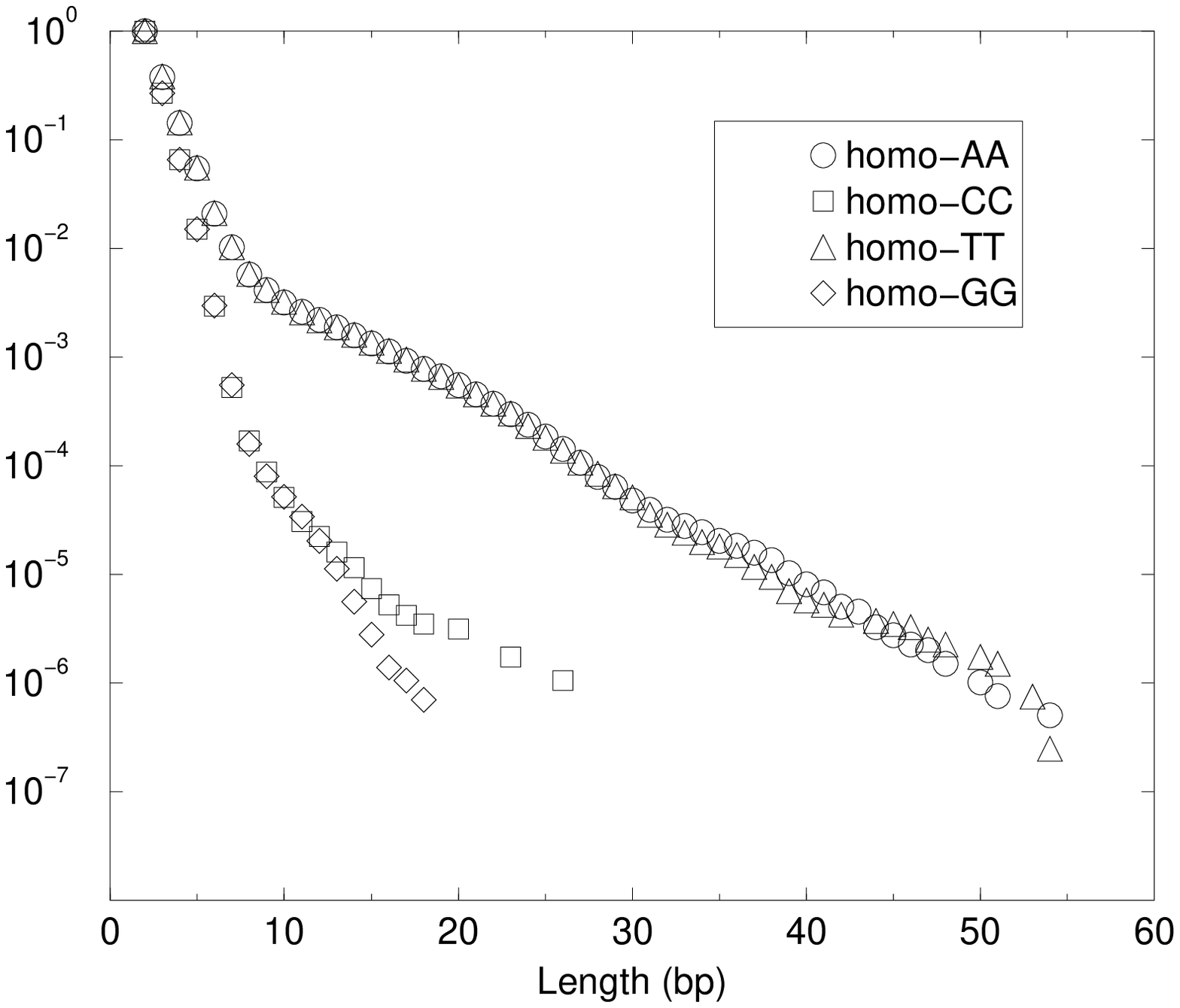}} 
\subfigure[]
{\includegraphics[width=6.7 truecm,clip]{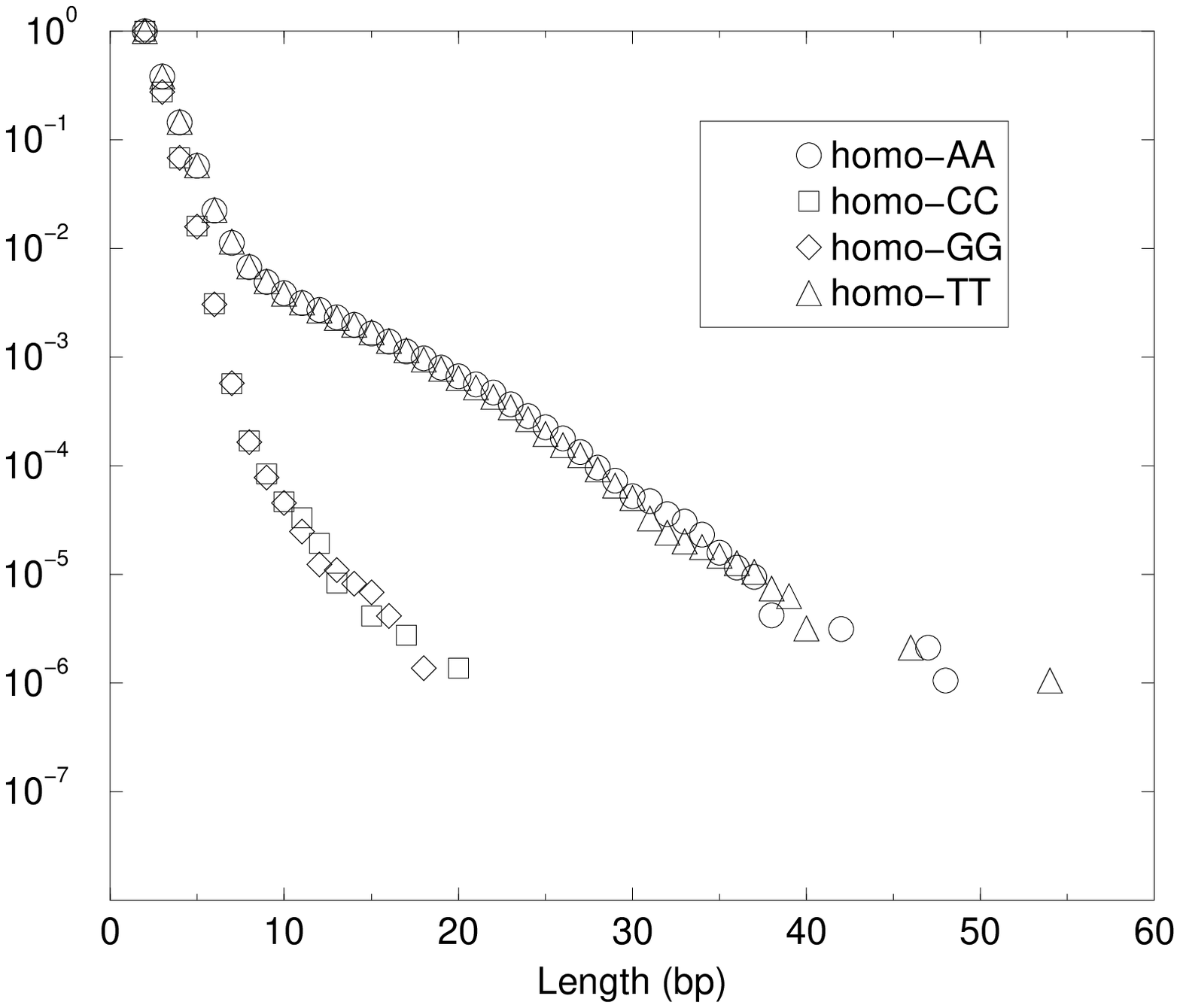}}
\caption{Cumulative length distributions of poly(X) words (X=[A,C,T,G]) in
non--coding regions in a representative case (Chromosome 1). (a) Intergenic
tracts. (b) Introns.}%
\label{fig_all.ncod}%
\end{figure}

\begin{figure}[ht]
\centering
\includegraphics[width=8 truecm]{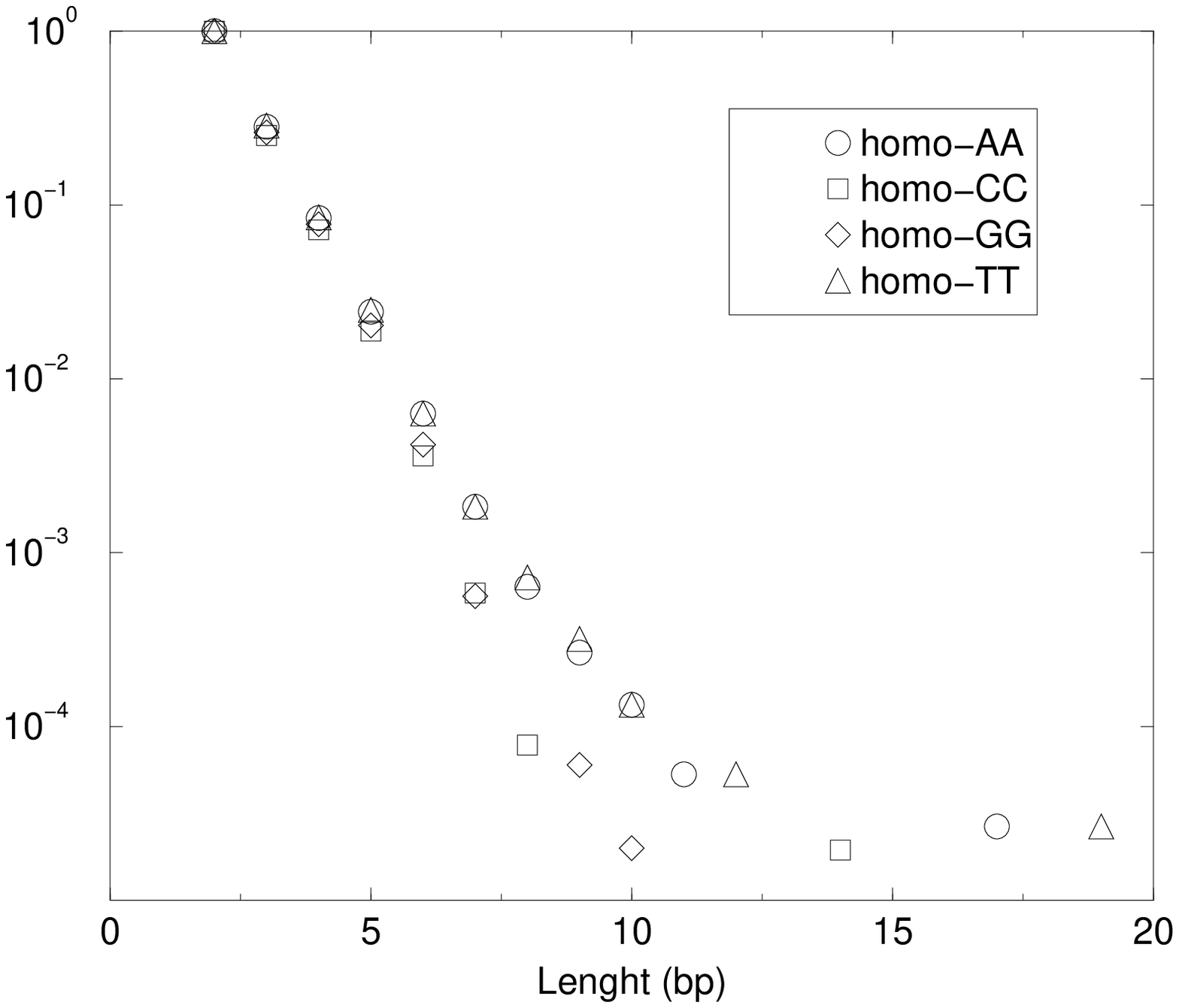}
\caption{Cumulative length distributions of poly(X) words (X=[A,C,T,G]) in
coding regions in a representative case (Chromosome 1).}
\label{fig_all.cod}%
\end{figure}

In non--coding regions there exists a clear difference between the
distributions of lengths of W--type words (poly(A) and poly(T)) S--type ones
(poly(C) and poly(G)).
%
Poly(A) and poly(T) words are much more represented and with much longer words
than poly(C) and poly(G) words (see Fig.~\ref{fig_all.ncod}). We do not
observe the same difference in coding regions (Fig.~\ref{fig_all.cod}). In
particular, the characteristic trends of such distributions in non--coding
regions are the same for inter--genic tracts and for introns.

\begin{figure}[htptb]
\centering
\includegraphics[width=8 truecm,clip]{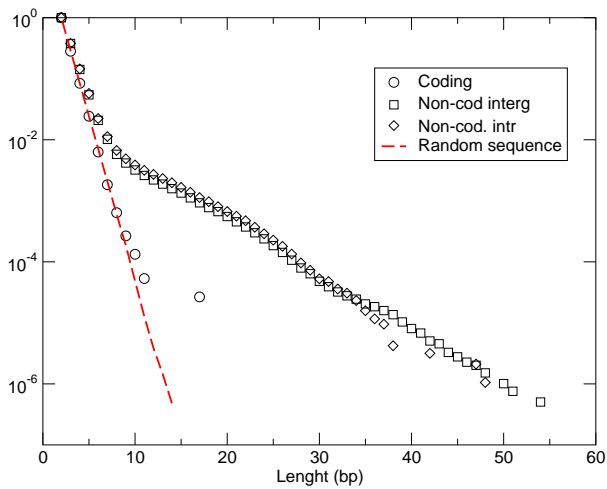}
\caption{{\small Comparison of the cumulative distributions of poly(A) tracts
in coding and non--coding regions with a simple Bernoulli model
(representative case of chromosome 1).}}
\label{homoAA}%
\end{figure}

In Figure~\ref{homoAA} we compare the cumulative length distributions of
poly(A) tracts in coding, non--coding inter--genic and intronic regions in a
representative case (chromosome 1, about $8$ \% of the genome) with the same
distribution from a fictitious sequence of the same length which has been
randomly reshuffled according to the same single--nucleotide probabilities and
coding/non--coding pattern. We note that the latter reshuffling reproduces in
the fictitious sequence the same set of exons as from the true sequence. It is
clear that non--coding DNA displays a marked over--abundance of poly(A) words,
whereas the length distribution found in coding tracts nicely fits within the
random uncorrelated scenario.

The distributions of all poly(X) sequences in all other human chromosomes
display the same statistical properties (data not shown).

In general, the peculiar behaviour of length distributions of poly(X) words in
non--coding regions are well described by a sum of two exponential laws, with
different length constants and weights. It is clear from the data shown in
Figure~\ref{homoAA} that the first exponential law describes a trend which is
to a good extent common to coding and non--coding regions. The second
exponential law takes over from a characteristic length onward. The latter
behaviour is very clear for poly(A) and poly(T) words, while it is less marked
for poly(C) and poly(G) words. This makes the identification of a crossover in
the length distributions of S--type words ambiguous. In the case of poly(A)
and poly(T) words, the crossover length can be calculated by separately
fitting the first and second portions of the length distributions. We call
$L_{0}$ the crossover length of the cumulative distributions 
$\mathcal{C}_{h}$:
\begin{equation}%
\begin{array}
[c]{lll}%
\mathcal{C}_{h}(L)=A_{1}\exp[\log(p_{1})L] & \text{for $L<L_{0}$}\nonumber\\
\mathcal{C}_{h}(L)=A_{2}\exp[\log(p_{2})L] & \text{for $L>L_{0}$} \quad .
\end{array}
\label{exp_law}
\end{equation}
Hence, we can calculate $L_{0}$ as
\begin{equation}
L_{0}=-\frac{\log(A_{1}/A_{2})}{\log(p_{1}/p_{2})}\quad.\label{Lzero}%
\end{equation}
The crossover length $l_{0}$ in the probability density will then be given by
\begin{equation}
l_{0}=L_{0}+\Delta l\quad,\label{l0}%
\end{equation}
where
\begin{equation}
\Delta l=-\frac{\log[\log(p_{1})/\log(p_{2})]}{\log(p_{1}/p_{2})}%
\quad.\label{Deltal}%
\end{equation}
The best--fit values of the parameters $A_{i}$ and $p_{i}$ ($i=1,2$) are
reported for chromosome 1 in Table~\ref{tab_fits} along with the corresponding values of
$L_{0}$ and $l_{0}$. We found similar results in the other chromosomes.
It is clear from the measured single--nucleotide and
di--nucleotide probabilities (see Fig.~\ref{f:pxx_vs_px}) that $p_{1}$ matches
to a good extent the single--nucleotide probabilities for poly(C) and poly(G)
words, while it corresponds to the di--nucleotide probabilities $p(A|A)$ and
$p(T|T)$ for poly(A) and poly(T) words. This means that in the first region
($L < l_{0}$) the length distribution of poly(X) words can be adequately
reproduced by treating words as uncorrelated or short--range correlated
sequences. This conclusion holds for both coding and non--coding regions. On
the contrary, it is apparent that the quantity $p_{2}$ cannot be associated
with any of the single-- or di--nucleotide probabilities, nor with any
$m$--nucleotide probability, with $m > 2$. This finding is consistent with the
general acceptance of the failure of zeroth-- and first--order Markovian
models of simple repeats to fully account for ``linguistic'' features of
non--coding DNA~\cite{Mantegna}.

Such intrinsic long--range effect is a signature of the peculiarities of DNA
replication dynamics. In particular, it is also indicative of a particular
mutual--help relationships between A--tracts and Alu repeats. The change in
slope of length distribution of poly(A) tracts reflects the characteristics of
DNA replication process. At a certain length of the repeats the replicative
process is more prone to errors. The poly(A) tracts elongate during DNA
replications by a slippage mutation mechanism, in such a way that the longer
the tract the more likely it is to change (elongate or
shrink)~\cite{ElongAslippage,Slippage2003}. The dynamical process explains
qualitatively the slope change in the length distribution of poly(A) tracts at
a specific length, which is the same in all chromosomes. Our estimates of such
characteristics length $l_{0} \approx10$ correlates well with the size of the
open complex of DNA during the replication (10 to 12 base pairs). On the other
hand, Alu elements multiply within the genome through RNA polymerase
III-derived transcripts in a process termed retroposition, thus contributing
with their poly(A) tails to the spreading of A--tracts through the genome.
This sort of mutual--help relationship between Alus and A--tracts is the major
finding reported in this paper.

In the next section, we shall establish in a clear quantitative way such link
between poly(A) tracts and Alus in the human genome.

\subsection{Statistics of separations between consecutive poly(X) words}

In this section we analyze the positioning patterns of poly(X) words along the
sequence of human chromosomes. Very generally, the statistics of separations
between consecutive words is a very useful tool in linguistics to isolate
content--bearing terms from generic ones. In general, the former words will
tend to cluster themselves as a consequence of their high specificity
(attraction or repulsion), while the latter ones will have a tendency to be
evenly distributed across the whole text. In order to eliminate the dependency
on frequency for different words, it is convenient to analyze the sequences of
normalized separations between consecutive words of length $L$,
$s=x(L)/\langle x(L) \rangle$. If homogeneous tracts were distributed at
random in the genome, the inter--tract distribution $P_{L}(s)$ for words of
length $L$ would be a Poissonian
\[
P_{L}(s) = e^{-s} \quad.
\]
As a consequence, we expect that non--specific words will run close to a
Poisson law, while larger deviations should occur for highly specific
content--bearing words. Such analysis may be implemented systematically in a
quantitative fashion by studying the standard deviations $\sigma_{L} =
\sqrt{\bar{s^{2}}-\bar{s}^{2}}$ of the distributions $P_{L}(s)$, which is the
simplest variable for characterizing a normalized distribution and its
fluctuations. For a Poisson distribution $\sigma_{L} = 1$, while if there is
attraction $\sigma_{L} > 1$. In the case of repulsion among words one should
expect $\sigma_{L}<1$.

\begin{figure}[th]
\centering
\includegraphics[width=12. truecm,clip]{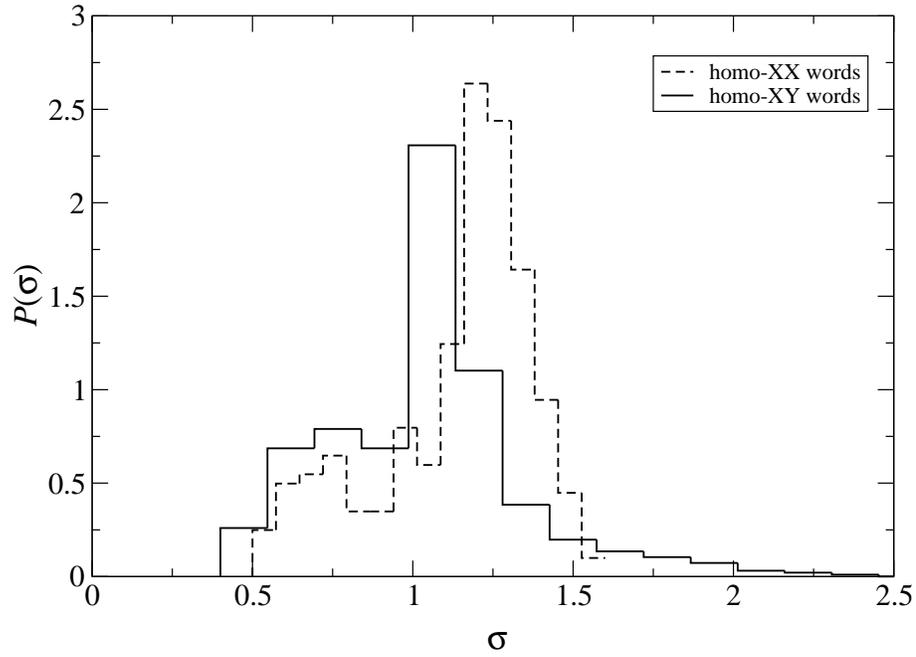}
\caption{{\small Normalized histograms of all standard deviations of the
distributions $P_{L}(s)$ for $L \geq2$ in the whole human genome. Comparison
of sequences of direct repeats of the kind poly(X) and repetitive sequences of
the kind poly(XY), coded according to the alphabets reported in
Table~\ref{t:alphabets}}}%
\label{fig:sigmahisto}%
\end{figure}

\begin{figure}[th]
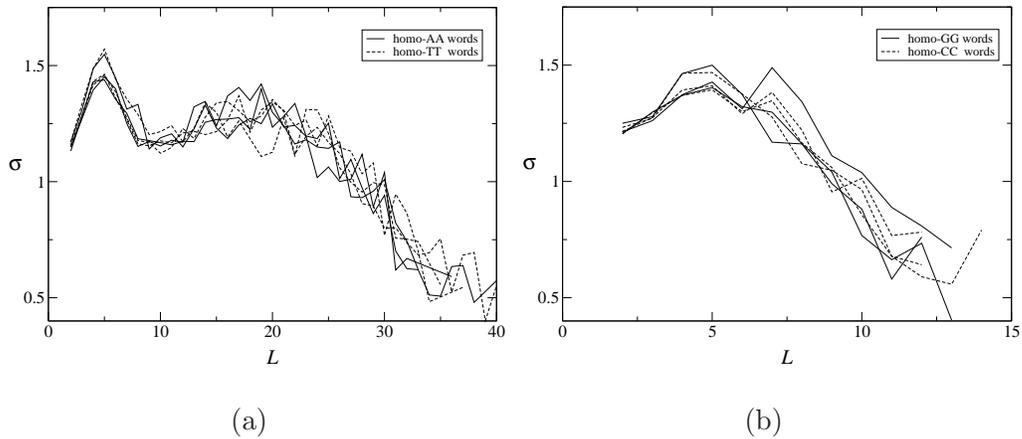

\centering
\subfigure[]
{\includegraphics[width=6.7 truecm,clip]{figures/Figure7.eps}}
\subfigure[]
{\includegraphics[width=6.7 truecm,clip]{figures/Figure8.eps}}
\caption{{\small Plots of the standard deviation of the distributions
$P_{L}(s)$ versus $L$ in a representative case (Chromosomes 1, 3, 5). (a)
poly(A) and poly(T) words, (b) poly(C) and poly(G) words}}
\label{fig:sigmavsL}%
\end{figure}

In Fig.~\ref{fig:sigmahisto} we compare the results of our analysis on
separations among tracts of direct repeats of given length in the whole human
genome with the result of the same analysis performed on short repeats of the
type poly(XY), coded according to the alphabets reported in
Table~\ref{t:alphabets}. The normalized histograms refer to distances between
HRs of lenght $L \geq2$. The figure clearly evidences that direct repeats are
more highly--specialized, higher--content words with respect to words coded in
the other alphabets, namely Purine/Pyrimidine (poly(AG) and poly(CT)) and
weak/strong (poly(AT) and poly(GC)). Interestingly, we find that the regions
where there is attraction ($\sigma_{L}>1$) or repulsion ($\sigma_{L}<1$) are
systematically associated with strings shorter or longer than a characteristic
length of about 25 bp, respectively. This effect is shown in a representative
case in Fig.~\ref{fig:sigmavsL}. We remark that we systematically obtain
analogous curves for all human chromosomes.

It is important to remark that the parameters $\sigma_{L}$ are estimated from
finite series of spacings. We may calculate the uncertainty $\Delta\sigma_{L}$
associated with such estimates in the hypothesis of random positioning of HRs.
In general, one has $\Delta\sigma_{L} = t_{\beta} \, [\mu_{4}/N_{L}
-(N_{L}-3)/N_{L}(N_{L}-1)\mu_{2}]$, 
where $N_{L}$ is the number of tracts of length $L$ in the series, 
$\mu_{n}$ is the $n$--th order moment, and $t_{\beta}$ specifies 
the required confidence level (e.g. $t_{\beta}=1$ for a confidence of one 
standard deviation, $\beta= 0.68$). 
Under the hypothesis of random positioning of HRs, we get 
$\mu_{4}=9, \mu_{2}=1$. Hence, at $68$ \% confidence, we have
\begin{equation}
\label{sigmaerr}
\Delta\sigma_{L}=\frac{8N_{L}-6}{N_{L}(N_{L}-1)}
\end{equation}
We have employed formula~(\ref{sigmaerr}) to check the statistical significance
of the values of $\sigma_{L}$ deduced from our series of spacings.
As an  example, for $L = 25$, we have in the worst case
$N_{25}=45, \sigma_{25}=1.05$ (chr. Y), and hence a relative error
$\Delta\sigma_{25}/\sigma_{25}\approx0.17$.  More generally, the number of tracts $N_{L}$
grows exponentially for $L<25$. This means that we rapidly get very small
errors on the estimates of $\sigma_{L}$ in the region of HR clustering. By the
same token, in the region $L>25$ we may still state that the observed
repulsion among tracts is statistically significative for nearly all
chromosomes, chromosomes smaller than the 20$^{\mathrm{th}}$ being at the
limit of statistical significance.

We interpret the clustering of poly(A) words shorter than about 25 bp in terms
of the correlation between Alu repeats and their flanking motifs of direct
repeats of A stretches~\cite{Batzer2002}. Our results are consistent with
recent finding by Holste \emph{et al.}, who show that histograms of distances
between adjacent Alu repeats show significant deviations from an exponential
decay, expected from random chromosomal positions of repeats~\cite{Holste2003}%
.
%
In other words, the content--bearing clustering of poly(A) oligomers shorter
than about 25 bp may be interpreted in terms of clustering of Alu repeats.
This conclusion establishes an important connection between the dynamics of
$A$ stretches and that of Alu repeats in the human genome.

\subsection{Distribution of separations in the $100-400$ bp region}

In order to further investigate the relation between the positions of poly(A)
sequences and of Alu repeats, we computed the distances between
poly(A)-poly(T) tracts considering only distances smaller than 800 bases. We
took that precaution in order to neglect other chromosome structures that may
be present on larger length scales and that may be subject of separate studies
in the future. For this purpose, we made use of kernel density plots to
analyze distance distribution. Histograms depend on the starting point on the
grid of bins and the differences between histograms realized with different
choices of the bin grid can be surprisingly large; another way to look at the
result is to compute the histograms averaging over a large number of shifts of
starting points and having very small bins~\cite{Scott1992}. Kernel density
estimators are smoother than histograms and converge faster to the true
density. The density function, which has unit total area, is computed through
the following formula
\begin{equation}
d(s)=\frac{1}{n}\sum_{j=1}^{n}\frac{1}{b}K\left(  \frac{s-s_{j}}{b}\right)
\quad,\label{densplots}%
\end{equation}
for a sample of distance values $s_{1},s_{2},\dots,s_{n}$, a given kernel
function $K(x)$ and a bandwidth parameter $b$. It can be shown theoretically
that the choice of the kernel is not crucial~\cite{wassermann}, whereas the
choice of the proper bandwidth is the important issue. The correct choice is a
compromise between smoothing enough to remove insignificant bumps and not
smoothing too much to smear real peaks away. We used a Gaussian kernel and we
selected data--dependent bandwidths, using the formula $b=0.9\,\mathrm{min}%
(\sigma,R/1.34)n^{-1/5}$, where $n$ is the sample size, $\sigma$ is the
standard deviation, and $R$ is the inter--quantile range~\cite{Silverman1986}.
Applying this formula, we determined the bandwidths for the human genome to be
$b=38.01$ bp. Bandwidths for different genomes resulted in very similar values.

\begin{figure}[th]
\centering
\includegraphics[width=15truecm]{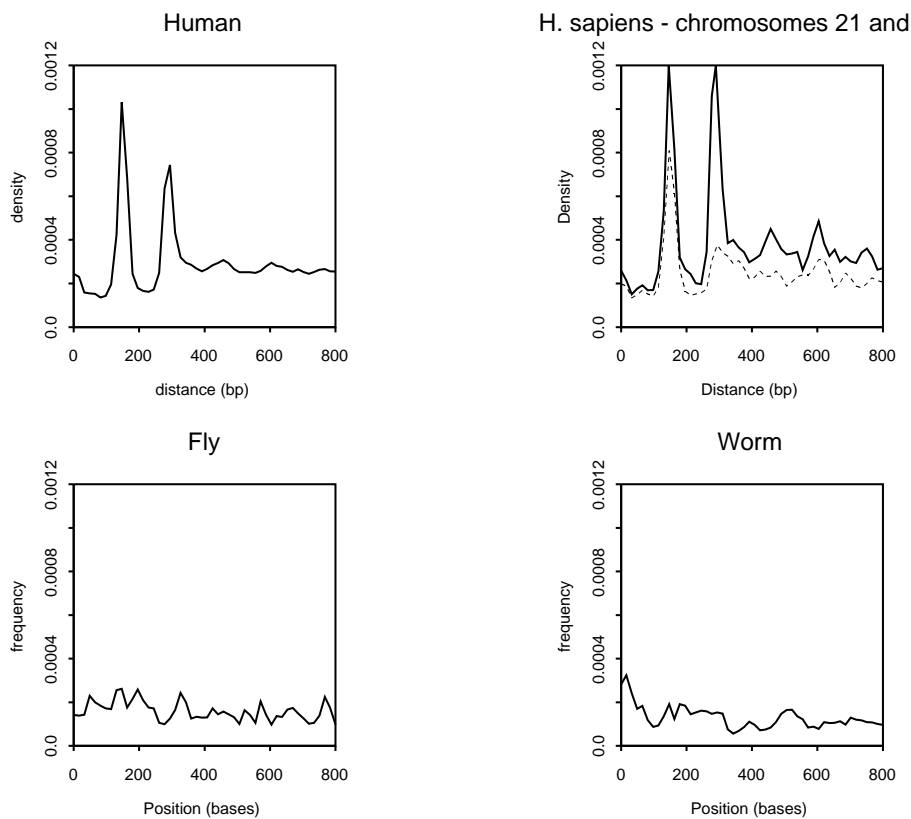}
\caption{{\small Distribution of poly(A) and poly(T) tracts in different
eukaryotes. Density plots of distances between poly(A) and poly(T) tracts of
12 bp (line) derived from (from top to bottom and left to right): 
(a) \emph{Homo Sapiens} entire genome; (b)
\emph{Homo Sapiens} chromosomes 21 (dotted line) and 22 (solid line); (c) the
\emph{Drosophila melanogaster} genome sequence (113.5 Mb) and (d) the
\emph{Caenorabditis elegans} genome sequence (87.6 Mb).}}
\label{f:DensPlot}
\end{figure}

In Fig.~\ref{f:DensPlot} we show the density plots of the distances between 12
bp poly(A)-poly(T) tracts in the human genome (a) and for chromosomes 21 and
22 (b). Many of the 12 bp tracts in the human genome sequences were found to
be positioned at intervals of either $\approx150$ or $\approx300$ bp. We found
similar patterns in all human chromosomes. Although chromosome 22 has the
highest tract density of the two (0.329 tracts/Kbp), its 300 bp band is very
pronounced and much bigger than that for chromosome 21 which has a much lower
tract density (0.217/Kbp). Using the chromosome 22 and 21 human sequences, we
also analyzed poly(A) and poly(T) tracts $>8$ bp in length, and also this
yielded distinct peaks at $\approx150$ and $\approx300$ bp (data not shown).
Other types of repeat sequences which are not thought to be
rigid~\cite{Bruckner1995,rigid,Bhattacharyya1999}, such as poly(AT) repeats of
12 bases, did not show this distinct periodicity. Next, we investigated
distances between longer poly(A)-poly(T) tracts and found that tracts longer
than 25 bases are located at longer and non periodical distances. This finding
is consistent with the general tendency of stretches of direct repeats longer
than about 25 bp to repel each other, which has been established in the
previous section. This might reflect the tendency of 
very long rigid homopolymeric tracts to be accomodated 
in the linker regions between nucleosomes so as to be 
as more scattered as possible
in order to favor the tightest packing of chromatine. 
Although in human chromosomes, the densities of
poly(A)-poly(T) tracts of 12 bases or more range from 0.188/Kbp (chromosomes
Y) to 0.374/Kbp (chromosome 16), there is in all a clear 150 bp periodicity,
and in most of the chromosomes the largest peak is the one at $\approx300$ bp.
The different intensities of the 150 and 300 bp periodicities in human
chromosomes reveal the clustering of Alus of 150 and 300 bp. The 150 bp
reflects the dimeric structure of Alus and shows that the central, short
poly(A)-poly(T) tract often elongates~\cite{Batzer2002}.

\begin{figure}[th]
\centering
\includegraphics[width=10truecm]{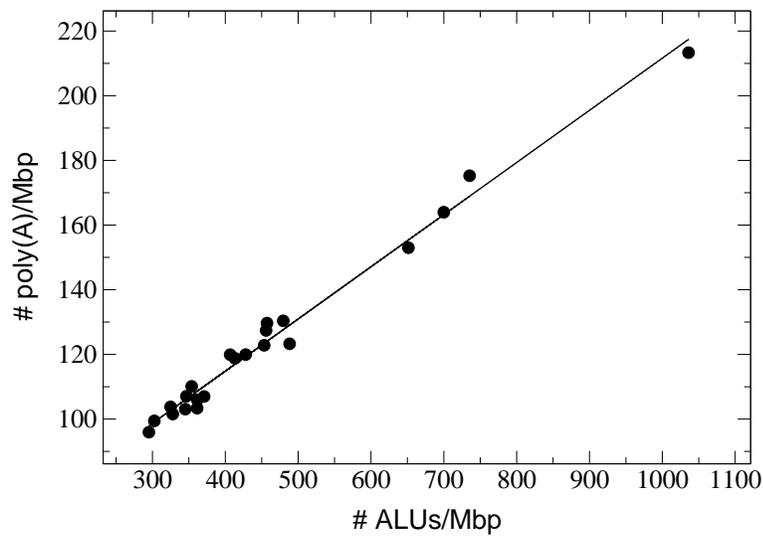}
\caption{Density of poly(A) tracts longer than 12 bp versus density of ALUs for the 24
human chromosomes (symbols) and linear fit (correlation coefficient
$r=0.993$)}%
\label{f:polyAvsALU}%
\end{figure}

In Fig.~\ref{f:polyAvsALU} we plot the density of poly(A) tracts longer than
12 bp (i.e. the number of tracts divided by the length of the chromosome)
versus the density of ALUs for each chromosome. The plot shows that poly(A)
longer than 12 bp are less abundant than Alus. Moreover, it is clear that a
linear correlation exists between the density of poly(A) and the density of
ALUs over the whole genome. If Alus were completely clustered, we would
observe 100 \% of the area under three peaks at 150, 300 and 450 bp. The
absence of the peak at 450 bp means that Alu clusters contain elements of the
same length (either 150 or 300 bp). We found that Alus are tandemly clustered
in small groups along each chromosome and that the different spacing between
consecutive clusters explains the peaks at 150 and 300 bp. Note that the
genomes of certain other eukaryotes such as \emph{C. elegans} (worm) and
\emph{D. melanogaster} (fly), that do not have ALU-like sequences, do not show
clear peaks and have much lower density of poly(A) (see Fig.~\ref{f:DensPlot}).

\section{Discussion}

The sequence of the human genome is highly repetitious at different sequence
length-scales and the coding sequences comprise less than 5\% of it. Many of
human genome repeats can be found in mature mRNA and total cellular RNA. RNAs
containing repetitive elements include Alu--containing mRNAs which amount to
5\% of all known mRNAs~\cite{Lander2001,WLI2}.

Patterns of homo and dinucleotide expansion in human genome suggest an
explanation as to why, contrary to vertebrate, low eukaryotes and bacteria
avoid the genome--wide accumulation and expansion of tandem repeats. Selection
acting on all of the repeats in a bacterial genome would generate a very high
mutational load (loss of many individuals due to selection) and would have to
act against very small incremental increases in genome size or repetitivity,
many of which would be expected to have minimal phenotypic effects. This high
cost of selection is tolerated in bacteria by the large population size and
the short cellular division time. The spreading of repeats across vertebrate
genomes occurred as non adaptive when the organism size increased and the
population-size decreased. In vertebrates, the selection against HRs may be
neutral or weakly negative because these regions are localised in the
periphery of the nucleus, where they are replicated
lately~\cite{EyreWalker2001,Fransz2002}. Moreover, even if they are inserted
in the regulatory or coding regions of genes, there is a high chance that
metabolic or genetic redundancy would buffer the effect.

\subsection{Why $A$s strings are more abundant than other strings?}

Strings of As and Ts have several peculiar properties not shared by strings of
Gs and Cs. They are very rigid, straight, show high stability with respect to
the mutation erosion and found in Alu repeats a perfect alliance for spreading
within the genome. First, in vitro studies have shown that such poly(A) and
poly(T) sequences can not be readily wound around the
nucleosome~\cite{Nelson1987}. Therefore, they remain exposed and not affected
by the silencing mechanisms~\cite{Iyer1995}. Since poly(A)-poly(T) tracts are
scarcely compatible with nucleosome formation, very long tracts may affect
chromatin organization. In humans, most of the large genome is organized in
gene--poor and densely compacted chromatin. This might involve a relatively
tight positioning of nucleosomes and thereby be responsible for the observed
spreading of Alus and, consequently, for the observed inter--tract statistics.

\subsection{The link between homopolymeric repeats and organismal complexity}

The abundance of HRs shows an apparent correlation with the organismal
complexity. For example, simple repeats are absent in viruses, rather rare in
bacteria and low eukaryotes and very abundant in high vertebrate genomes. A
strong association has been found between organismal complexity and the
complexity of regulatory regions upstream and downstream the genes and the
complexity of the coding region.

In a typical mammalian gene, HRs and other repeats or mobile genetic elements
can have different effects on gene functioning. In particular, HRs can be found
in both the regulatory and the coding regions. The multiple regulatory regions
upstream the gene are binding sites for transcription factors and represent
subfunctions that might finely tune, positively or negatively, the level of
expression in a specific tissue and developmental stage. The coding region
generally codes for proteins with several structural/functional domains that
may interact with different ligands/proteins.

Phenotype and the acquisition of new gene functions has been associated with
gene duplication~\cite{Lynch2000,Wagner1998}. We hypothesize that it can be
also affected by the presence of HRs. The probability of preservation of both
gene duplicates increases with the number of independent subfunctions in the
regulatory or coding region due to a greater number of ways that gene
duplicates can differently evolve, by means of keeping or loosing some of
these subfunctions. The insertion of repeats, mobile elements and the
elongation of homo or dimer strings in the regulatory region in one duplicate,
may change the binding affinities between transcription factors and the basal
transcription machinery and increase the probability of recombination, loss,
acquisition, shuffling, duplication of regulatory sites with respect to the
other gene copy. The same may occur in the coding regions. These events will
increase rather than reduce the probability of duplicated gene fixation
because each gene can now perform a function the other gene cannot, for
example the two genes being expressed in different tissues or developmental
stage~\cite{Wagner1998,Dermitazakis2001,Lynch2001,Wagner2001,LynchForce2000,Force1999}%
. Moreover, several hundred genes use fragments of mobile elements in the
regulatory sequences that control expression and transcription
termination~\cite{Deininger2003,Batzer2002}. This suggests that, at least in
part, mobile elements such Alus are retained because they confer some
advantages~\cite{Han2004}. SINEs and LINEs appear to be subject to RNA
interference (RNAi) that is a form of post-transcriptional gene silencing
triggered by double-stranded RNA~\cite{Deininger2003}.

The probability of fixation by differential subfunctionalization approaches
zero in large populations because the long time to fixation magnifies the
chances that secondary mutations will completely incapacitate one copy before
joint preservation is complete; therefore, subfunctionalization is a more
important factor in high eukaryotes than in bacteria and low eukaryotes, where
neofunctionalization, i.e. arising of completely new genes, is a more frequent event.

Several examples of subfunctionalization of regulatory and coding regions are
reported in literature~\cite{LynchForce2000}. We hypothesize that HRs affect
the mobility of Alus and increase in a subtle way the combination and
diversification of gene expression and the fixation of multigene families.
Since poly(A)-poly(T) tracts are scarcely compatible with nucleosome
formation, and very long tracts may affect chromatin higher structure, we aim
at investigating how their distribution can be used as an indirect means to
obtain insights into the structural organization of DNA, on both genome-wide
scale and on individual chromosomes.

\section{Conclusions}

Despite the availability of several high eukaryote genomes, the evolutionary
dynamics of the simplest repeats are not yet fully understood. Since
microsatellite slippage mutation rates depend on many factors, among which,
repeat motif-length, here, we have studied the genome-wide base composition of
the microsatellites and we have particularly focused on the relationships
between poly(A) and Alus in the human genome.

We have shown by means of standard linguistic analysis that HRs are
highly--specific, content--bearing terms within the DNA sequence of humans.
More specifically, we have provided evidence that a quantitative analysis of
length and inter--tract distributions of HRs may provide insight into the
mobility of these elements within the genome.

The clear 150 and 300 bp periodical patterns of poly(A)-poly(T) tracts,
revealed with the aid of a kernel density estimator, show that these tracts
are almost entirely associated to Alus and to mobile elements of similar
length. Since most Alus are 300 bp long, the fact that the signal at 150 bp is
higher than the one at 300 bp in almost all distributions of separations
between A--tracts and T--tracts from all chromosomes suggests that the
central, short poly(A)-poly(T) tracts often elongates.

We have shown that a quantitative analysis of the link between poly(A) repeats
and Alus has important consequences on the understanding of the joint dynamics
of Alus and simple repeats in the human genome. On one hand, the intrinsic
rigidity of sequences such as poly(A) tracts helps Alus mobilization,
hindering packing of Alu--containing tracts into the nucleosome structure. On
the other hand, middle and 3' end regions of Alus are source of longer
poly(A)-poly(T) tracts. Furthermore, these tracts may mutate and populate also
other poly(X) sequences. Work in progress focuses on reconstructing the
dynamics of Alu clusters formation in the human genome. In conclusion, we wish
to stress that the genome-wide statistical analysis of low complexity is a
thriving field of research. In particular, it may have two important benefits:
improve the understanding of the processes that shaped the genome organization
and improve the ability to correlate phenotype complexity with genome organization.

\section{Acknowledgements}

F. P. acknowledges funding from the Italian Institute for Condensed Matter
Physics, under the Forum project STADYBIS.

%

\newpage

\centerline{\bf Tables}

\vspace{2truecm}

%
\begin{table}[th]
\centering%
\begin{tabular}
[c]{c|cccc}\hline
alphabet & $A$ & $C$ & $G$ & $T$\\\hline\hline
S/W & W & S & S & W\\
R/Y & R & Y & R & Y\\\hline
\end{tabular}
\caption{List of different alphabets used to code simple repeat stretches. W
(weak) and S (strong) code with respect to the strength of the inter-strand
H--bond between pairs of complementary nucleotides. R (purine), Y
(pyrimidine).}%
\label{t:alphabets}%
\end{table}
%

\vspace{2truecm}

%
{\small \begin{table}[h]
{\small \centering%
\begin{tabular}
[c]{c|cccc|cc}\hline\hline
\textbf{poly(A)} & $A_{1}$ & $p_{1}$ & $A_{2}$ & $p_{2}$ & $L_{0}$ (bp) &
$l_{0}$ (bp)\\\hline\hline
\emph{coding} & 12.7 & 0.28 & - & - & - & -\\
\emph{non--cod.} & 6.09 & 0.39 & 0.029 & 0.81 & 7.4 & 9.45\\\hline\hline
\textbf{poly(T)} & $A_{1}$ & $p_{1}$ & $A_{2}$ & $p_{2}$ & $L_{0}$ (bp) &
$l_{0}$ (bp)\\\hline\hline
\emph{coding} & 12.8 & 0.28 & - & - & - & -\\
\emph{non--cod.} & 6.09 & 0.39 & 0.026 & 0.81 & 7.6 & 9.65\\\hline\hline
\textbf{poly(C)} & $A_{1}$ & $p_{1}$ & $A_{2}$ & $p_{2}$ & $L_{0}$ (bp) &
$l_{0}$ (bp)\\\hline\hline
\emph{coding} & 21.85 & 0.23 & - & - & - & -\\
\emph{non--cod. interg.} & 24.0 & 0.22 & 2.5 $\times10^{-4}$ & 0.8 & 8.9 &
10.4\\
\emph{non--cod. intron.} & 23.5 & 0.22 & 1.4 $\times10^{-3}$ & 0.7 & 8.4 &
9.65\\\hline\hline
\textbf{poly(G)} & $A_{1}$ & $p_{1}$ & $A_{2}$ & $p_{2}$ & $L_{0}$ (bp) &
$l_{0}$ (bp)\\\hline\hline
\emph{coding} & 22.6 & 0.23 & - & - & - & -\\
\emph{non--cod. interg.} & 23.4 & 0.22 & 0.014 & 0.57 & 7.8 & 8.84\\
\emph{non--cod. intron.} & 23.3 & 0.22 & 1.95 $\times10^{-3}$ & 0.67 & 8.4 &
9.6\\\hline\hline
\end{tabular}
}\caption{Results of the fits performed with the function~(\ref{exp_law}) on
the poly(X) length distributions in a representative case (chromosome 1). Note
that no appreciable difference is found in the best--fit values of the
floating parameters between intergenic and intronic regions for the
distributions of poly(A) and poly(T) words.}%
\label{tab_fits}%
\end{table}}

\newpage

\centerline{\bf Figures}

\vspace{2truecm}

%
%

%
%
%

%
%

%
%

%
%

%
%
\end{document}